\documentstyle[prd,preprint,aps,epsfig]{revtex}

\begin{document}


\newcommand{\TeV}{\,{\rm TeV}}
\newcommand{\GeV}{\,{\rm GeV}}
\newcommand{\MeV}{\,{\rm MeV}}
\newcommand{\keV}{\,{\rm keV}}
\newcommand{\eV}{\,{\rm eV}}
\def\ap{\approx}
\def\bea{\begin{eqnarray}}
\def\eea{\end{eqnarray}}
\def\bi{\begin{itemize}}
\def\ei{\end{itemize}}
\def\be{\begin{enumerate}}
\def\ee{\end{enumerate}}
\def\ler{\lesssim}
\def\gtr{\gtrsim}
\def\beq{\begin{equation}}
\def\eeq{\end{equation}}
\def\haf{\frac{1}{2}}
\def\nn{\nonumber}
\def\p{\prime}
\def\ccg{\cal G}
\def\L{\cal L}
\def\O{\cal O}
\def\R{\cal R}
\def\U{\cal U}
\def\V{\cal V}
\def\W{\cal W}
\def\e{\varepsilon}
\def\slash#1{#1\!\!\!\!\!/}

\setcounter{page}{1}
\draft
\preprint{KAIST-TH 2000/14, hep-th/0012091}

\title{\Large \bf A criterion for admissible singularities in brane world}

\author{Hyung Do Kim}

\address{Department of Physics,
Korea Advanced Institute of Science and Technology\\
        Taejon 305-701, Korea \\
{\tt hdkim@muon.kaist.ac.kr}}


\tighten

\maketitle

\begin{abstract}
When gravity couples to scalar fields
in Anti-de Sitter space,
the geometry becomes non-AdS and develops singularities generally.
We propose a criterion that
the singularity is physically admissible
if the integral of the on-shell Lagrangian density over
the finite range is finite everywhere.
For all classes of the singularities studied here,
the criterion suggested in this paper 
coincides with an independent proposal made by Gubser
that the potential should be {\it bounded from above} in the solution.
This gives a reason why Gubser's conjecture works. 
\end{abstract}

\newpage

\section{Introduction}
The brane world scenario has been extensively studied
for last several years.
The discovery of D-brane \cite{polchinski}
in string theory changed the conventional
viewpoint that the spacetime on which gauge fields propagate
and gravitons propagate should be the same.
The gauge fields can be confined on D-brane;
the fact that gauge fields and gravitons can propagate
in different spacetimes allowed us to look at 
traditional problems in entirely new ways.
The gauge hierarchy problem, which has been a central issue in particle physics
for several decades,
is now posed differently
as why the Planck scale is so high (gravity is so weak)
compared to the electroweak scale (gauge interactions)
\cite{ADD,RS1,RS2}.
A large extra dimension \cite{ADD} diluting gravity explains
the gauge hierarchy if a natural mechanism
to stabilize the radion at large values
can be found.
Alternatively,
Randall and Sundrum \cite{RS1} proposed that if the extra dimension
is negatively curved (Anti-de Sitter) and 
if we live on the negative tension brane,
we can explain the huge discrepancy between the Planck scale
and the electroweak scale without invoking large extra dimension.
As an extension of this work, they also showed that the gravity
can be confined \cite{RS2} even for the extra dimension of infinite size
if the tension of the brane and the bulk cosmological
constant has a special relation.

While all the above scenarios have been invented to explain the gauge hierarchy
problem, the cosmological constant problem still remains a serious conundrum. 
If the fine tuning between the bulk cosmological constant
and the brane tension is incomplete,
we get effective 4 dimensional cosmological constant proportional
to the mismatch \cite{Nihei,Kaloper,KimKim,Steinhardt}.
In this paper we provide no further insights to this issue
and simply assume the fine tuning allowing us to restrict our attention
only to the space-time geometries with 4 dimensional Poincare invariance.

The generalization of Randall-Sundrum (RS) setup
has been studied extensively. For example, see \cite{G1,G2}.
The simplest interesting extension of RS setup is 
to consider bulk scalar fields coupled to gravity.
First of all, known string theories contain various scalar fields
such as dilaton, moduli and axions.
As a first step toward the stringy generalization of RS scenario,
the inclusion of the bulk scalar fields appears unavoidable.
Furthermore, in two brane scenario \cite{RS1}, 
bulk scalar field is necessary for the radion stabilization \cite{GW,DFGK};
without stabilization mechanism, pure gravity reveals the instability
of the two brane systems \cite{KimKim}.
One also notes that the bulk cosmological constant 
can be naturally generalized to the potential of scalar fields.
In this paper we consider a gravity coupled to a single bulk scalar field.

According to the AdS/CFT correspondence \cite{AC1,AC2,AC3},
supergravity in $AdS_{D+1}$ is dual to $D$ dimensional
conformal field theory.
Extensions of the AdS/CFT correspondence 
to the duality between supergravities in non$AdS$ backgrounds
and the field theories off the conformal fixed points have been suggested
in supersymmetric context
\cite{Girardello,Freedman,johnson,Pol2,Klebanov,MN1,MN2,johnson2}.
One of the general properties of these models 
is the appearance of naked singularities in the IR region
of the background geometry.
A concrete explanation for the appearance of the singularity
can be found in \cite{cosmoKim}.
The explanation is based on the analogy with inflation.
Most inflationary scenario has a scalar field, inflaton,
whose potential has a minimum or degenerate minima
with vanishing vacuum energy implying that
the potential is globally nonnegative.
Though there are two types of solutions (inflating and ``deflating")
due to time reversal symmetry in the Einstein equation,
we are interested in inflating solutions in which
the affine connection terms act as a frictional force.
With the aid of the friction, the inflaton settles down
to a minimum. The minimum of the potential is always an attractor
of the system in inflating models 
and we recover asymptotically de Sitter geometry.
Just as we consider the motion of inflaton along the time direction,
we consider the change of the bulk scalar field
along the $y$ (extra dimension) direction.
Starting from an initial position $y=0$,
the scalar field develops a $y$-dependent profile 
determined by the equations of motion.
Since we are interested in the brane world scenario
whose $D$ dimensional Newton constant
is finite, only the decreasing warp factor should be considered.
The following discussions are mainly independent
of the presence of the branes and can be applied to whole AdS
geometry in the same way.
Now the situation is analogous to that of the ``deflating" solution 
or time-reversed inflation.
The crucial difference is that
the affine connection terms act as an {\it anti-friction}.
They prevent the scalars from settling down 
at a stationary point of the potential
except when the initial condition has been precisely chosen to do it.
The generic final destiny of the scalar is to roll up or down to infinity,
producing a singularity.
If this happens at finite $y=y_c$, 
there is a naked singularity.

These singularities appear in generic situations.
Here, we propose a criterion that determines
which types of the singularities are physically acceptable.
\begin{quote}
\it
If the integral of the on-shell Lagrangian density over the finite 
range of $y$, whose least upper bound is $y= y_c$, is finite, the 
singularity at $y=y_c$ is physically admissible. 
\end{quote}
In \cite{gubser}, a different version of the criterion
on the physically acceptable singularities was given.
\begin{quote}
\it
Large curvatures in scalar coupled gravity with
four dimensional Poincare invariant solution
are allowed only if the scalar potential is bounded above
in the solution.
\end{quote}
The main observation of this paper is that the above two criteria
that apparently look independent are equivalent
for large class of known interesting examples.
We expect that the two criteria are equivalent in general cases,
even if we do not have a rigorous mathematical proof yet.
The conjecture in \cite{gubser} was based on
the detailed study of the known supergravity examples.
For the model having the potential unbounded from above,
it has been shown that we always encounter 
a pathological problem when we see the singularity
in the dual field theory or when we want to resolve
it by lifting it to higher dimensions in string theory.

There are other criteria on the singularities 
\cite{gellmann,cohen,minic,MN1}.
In \cite{gellmann,cohen}, the unitarity condition at the singularity
is used to probe which types of singularities are harmless.
In \cite{MN1}, the $g_{00}$ component of the metric is required
to be bounded above (or not to increase as we approach the singularity)
for physically allowed singularities.
It is likely that all these criteria share the common features.
However, the connections among them are not clear,
and to show the detailed connection is not discussed in this paper.

The rationale for our criterion
comes from two sources.

First, to have a sensible semi-classical expansion
around a given classical solution, 
it is necessary that the integral of the on-shell Lagrangian 
density over any finite volume be finite \cite{coleman}.
String theory in curved spacetime is far from complete.
Curved spacetime is treated as the background,
and at best we can take the semi-classical expansion around the background
as far as the gravity sector is concerned.
In the absence of full quantum description of gravity,
the only way on which we can rely is to use the semi-classical expansion.
This approach is based on the belief that 
the semi-classical expansion grasps most of important physics.
In other words the belief is that
the difference between the fully quantized theory
and the semi-classical one is negligible
and the semi-classical treatment is trustworthy.
However, once semi-classical expansion is not available,
we can not trust the classical solution 
since the fully quantized theory is expected to be dramatically different
from the classical one.
The information gained from the geometry containing harmful singularities,
which do not allow the semi-classical expansion,
is therefore not trustable, and we expect the quantum effects
of gravity will spoil the picture completely.
In that case we should abandon classical general relativity description.

Second, to satisfy the consistency condition
with putting a finite tension brane at the singularity,
the finiteness of the on-shell Lagrangian density
from $y=0$ to $y=y_c$ is also required.
There are several consistency conditions in the brane world scenario
which become important in the presence of the singularity
\cite{Ellwanger,Nilles,Kallosh}.
For the metric which keeps $D$ dimensional Poincare
invariance in $D+1$ dimension, the consistency
requires that the $D$ dimensional energy density
(or effective $D$ dimensional cosmological constant)
after integrating out the extra dimension $y$ should vanish.
Already in self-tuning model \cite{self1,self2}
it has been shown that we
need an additional contribution to the effective 4 dimensional cosmological
constant from the singularity to cancel the brane tension.
In self-tuning model, the bulk cosmological constant is assumed to be zero
and there is no other contribution to the 4 dimensional cosmological constant
except the brane tension.
This inconsistency problem can be overcome 
by putting additional brane at the position
of the singularity such that the tension of it can cancel the tension of 
the visible brane at $y=0$. 
In general, if the 4 dimensional energy density
obtained by integrating over $y$ from $y=0$ to $y=y_c$
is finite, the consistency condition can be satisfied by putting
additional brane on the singularity,
and it leaves open a possibility 
that the singularity can be resolved in some ways.
However, if the energy density integrated over $y$ including the singularity
is infinite, we can not make a consistent theory without introducing
an artificial brane with infinite tension located at the singularity.
This implies that the physical system, which will be obtained only after
resolving the singularity, is entirely different from the classical
configuration.

The criterion proposed here has nothing to do with supersymmetry.
While all the known examples which have concrete realizations
are supergravity models, our criterion is 
more fundamental
and is also applicable to non-supersymmetric cases.

\section{Basic setup}
We start from D+1 dimensional geometry with D Poincare invariant spacetime.
The metric is
\bea
\label{metric}
ds^2 & = & a^2(y) \eta_{\mu\nu} dx^{\mu} dx^{\nu} + dy^2,
\eea
and we define ``Hubble parameter" as
\bea
H & = & \frac{a^\p}{a}  = -h,
\eea
where the prime denotes the derivative with respect to $y$.
This new ``Hubble parameter" is defined to use an analogy
with the inflation models where the derivatives are taken with respect to $t$
instead of $y$.

Since we are interested in the geometry with decreasing warp factor
such that $D$ dimensional Newton constant can remain finite,
$H$ is always negative definite, and $h$ is positive definite.($h \ge 0$)
In discussing the properties of the singularities,
other parts cut out by the brane
which approach Anti de Sitter space asymptotically
do not affect the conclusion \cite{gubser}.
Once we start from negative $H$ and go to the direction
of decreasing warp factor, the holographic c-theorem
guarantees that $H$ remain negative definite.
{\footnote{The absolute value $h$ increases monotonically unless we introduce
an object that violates the null energy condition.}}

D+1 dimensional action of scalar coupled gravity is
\bea
S & = & \int d^{D+1}x \sqrt{-g} (\frac{1}{2} R 
- \frac{1}{2} \partial_{\mu} \phi 
\partial^{\mu} \phi - V(\phi) ) \\ \nn
&&+ \int d^Dx \sqrt{-g^{(D)}} {\cal L}_i,
\eea
with the unit of setting the fundamental scale $M_{D+1} = 1$.
Single real scalar field is considered in this paper,
but the generalization to the system including many scalars
does not affect the result given in the paper
since the singular behavior can be effectively described
by single real scalar field.
Let us assume ${\cal L}_0 = -V_0$ from now on
and the brane is located at $y= y_0 = 0$.
The brane is introduced in order to see the finiteness of the on-shell
Lagrangian density more clearly. $D$ dimensional energy density
is well defined in the presence of the brane cutting
the boundary part of AdS (asymptotically AdS) which
is irrelevant in the discussion of the physics about the singularities.
Einstein equations are summarized as two equations 
for the metric in eq. (\ref{metric}) 
\bea
&&\phi^{\p\p} + D H \phi^\p - \frac{dV}{d\phi} = 0, \\ \nn
&&H  = - \sqrt{\frac{2}{D(D-1)}\left(\frac{\phi^{\p 2}}{2}-V\right)}.
\eea

It is helpful to introduce new notation $U = -V$ with $h= -H$
since the new quantities $U$ and $h$ are mainly positive quantities.
The Einstein equations are 
\bea
&&\phi^{\p\p} - D h \phi^\p + \frac{dU}{d\phi} = 0, \\ \nn
&&h = \sqrt{\frac{2}{D(D-1)}\left(\frac{\phi^{\p 2}}{2}+U\right)}.
\eea
It is clear that these two equations are exactly the same
as the inflaton equations
with Hubble parameter $-h$ and potential $U$
(or $y$ to $t$).
Since the Hubble parameter $-h$ is negative,
we should think of it as "deflation" or
time-reversal solution to the usual inflation motion \cite{cosmoKim}.
The comparison with the inflation models are well summarized
in the table.

The equation looks simpler if we set
$h = \sqrt{\frac{2}{D(D-1)}~G}$.
We now have
\bea
&&G = \frac{\phi^{\p 2}}{2} + U , \\ \nn
&&G^{\p} = (\phi^{\p\p} + \frac{dU}{d\phi}) \phi^{\p},
\eea
and the ``Friedmann equation" becomes
\bea
\label{G}
&&G^{\p} - \sqrt{\frac{2D}{(D-1)}~G}~ \phi^{\p 2} = 0, \\ 
\label{U}
&&\phi^{\p 2} = 2 (G-U).
\eea
We can rewrite the ``Friedmann equation" by eliminating $\phi^\p$
\bea
\label{friedmann}
&&G^{\p} - 2 \sqrt{\frac{2D}{(D-1)}~G}~ (G-U) = 0.
\eea

Furthermore, for the classical solution of the Einstein equation,
\bea
R & = & \partial_{\mu} \phi \partial^{\mu} \phi
+ \frac{2(D+1)}{D-1} V + \frac{2D}{D-1} V_0 \delta(y),
\eea
and if we put it again into the original action, the action is expressed
only in terms of $V$ independently of $\phi^{\p}$.
\bea
\label{energy}
S & = & \int d^{D+1}x \sqrt{-g} ( \frac{2}{D-1} V ) 
    + \int d^Dx \sqrt{-g^{(D)}} ( \frac{1}{D-1} V_0 ).
\eea

A particular boundary condition on the brane at $y=0$ should be considered 
to have a consistent $D$ dimensional Poincare invariant solution.
The boundary jump condition is
\bea
H |_{-\epsilon}^{\epsilon} & = & - \frac{1}{D-1} V_0 ( \phi (y=0)) \\ 
\label{b1}
\phi^{\p}|_{-\epsilon}^{\epsilon} & = & 
\frac{\partial V_0 (\phi(y=0))}{\partial \phi}.
\eea
For positive definite $h$,
the boundary jump condition is
\bea
\label{b2}
h |_{-\epsilon}^{\epsilon} & = & \frac{1}{D-1} V_0 ( \phi (y=0)) .
\eea

In the following analysis, $V_0$ is not specified
and is assumed to be chosen to satisfy the junction condition
of eq. (\ref{b1}) and (\ref{b2})
once we determine $\phi^{\p}$ and $h$ at $y=0$
(equivalently $G$ or $\sqrt{G}$ at $y=0$).
The boundary jump condition is necessary to give $D$ dimensional Poincare
invariant solution.
Otherwise, we can not have a solution
for the metric of eq. (\ref{metric})
with $D$ dimensional Poincare invariance
and end up obtaining different solutions, namely, $dS_{D-1}$ or $AdS_{D-1}$
\cite{Nihei,Kaloper,KimKim,Steinhardt}.

\section{Case studies}
\subsection{U$=$ constant}

Suppose the potential $-U$ is constant and negative
($U = U_0$ is positive).
From the eq. (\ref{friedmann}), we have
\bea
\frac{dG}{\sqrt{G}{(G-U)}} = \sqrt{\frac{8D}{(D-1)}} dy.
\eea
The solution to the above equation is
\bea
\sqrt{G} = \sqrt{U_0} \coth {\left(\sqrt{\frac{2D}{D-1}}(y_c-y)\right)}.
\eea
We are interested in $G \ge U_0$ and thus $y \le y_c$.
In this case, $\sqrt{G}$ goes to infinity as $y$ approaches $y_c$.
Now $\phi$ can be calculated from $U_0$ and $G$.
\bea
\phi^{\p 2} & = & 2(G-U) = 
\frac{2U_0}{\sinh^2 {\left(\sqrt{\frac{2D}{D-1}}(y_c-y)\right)}} , \\ \nn
\phi^{\p} & = &
\frac{\pm \sqrt{2U_0}}{|\sinh {\left(\sqrt{\frac{2D}{D-1}}(y_c-y)\right)|}}. 
\eea
\bea
\phi = \phi_0 \pm \sqrt{2U_0} \log \tanh {\left(\sqrt{\frac{2D}{D-1}}(y_c-y)\right)}.
\eea

The ``$D$ dimensional energy density" is
defined by putting the scalar curvature
$R$ calculated from equations of motion into the Lagrangian
and integrating over $y$ (\ref{energy}).
The next thing we can do is to check whether 
the ``$D$ dimensional energy density"
is finite or not.
The precise $D$ dimensional energy density is
\bea
I_D & = & \frac{2}{D-1} I
+ \frac{1}{D-1} V_0 \nn \\
S & = & \int d^Dx I_D. \nn
\eea
where
\bea
I & = & \int_0^{y_c} dy \sqrt{-g} V.
\eea
Since we are only interested in discriminating
whether the "$D$ dimensional energy density" diverges or not,
the numerical coefficients, the sign, 
and the finite contribution from the brane tension
are neglected in the following discussions.
Whether the quantity $|I|$ converges or diverges is the only important question
in this paper.
In the first example, the ``$D$ dim. energy density" $|I|$ 
is
\bea
|I| & = & \int_0^{y_c} dy e^{-\int_0^y d\bar{y} ~\sqrt{\frac{2D}{(D-1)}~G}} ~U,
\eea
and is finite.
Therefore, the singularity developed in the flat potential
satisfies the criterion,
and we conclude that this singularity is admissible.
The singularity appearing in the self-tuning model \cite{self1,self2}
belongs to this case.
It is lucky if the exact solution is available which is the case here.
In most cases it is not possible to obtain an analytic solution
which is valid in all ranges from $y=0$ to $y=y_c$.
Nonetheless, it does not prevent us from checking
whether some classes of the singularities satisfy
the criterion or not once we know the behavior of the solutions
near the singularity.
The finiteness or infinity of the $D$ dimensional energy density
is determined only by the behavior in the neighborhood of the singularity.
Thus in the following examples, we classify the solution
by the singular behavior
of the bulk scalar field and/or the metric
and reconstruct the leading term of the potential 
when we approach the singularity.

As $y \rightarrow y_c$, the singular behavior can be characterized by
\bea
\label{limit}
\phi^{\p} = \frac{A}{(y_c - y)^{\alpha}}
\eea
where $A$ and $\alpha$ are arbitrary numbers.
The limiting behavior of $\phi$ is then
\bea
\phi = \frac{A}{\alpha-1}\frac{1}{(y_c-y)^{\alpha-1}}.
\eea
From the eq. (\ref{friedmann}), we get
\bea
\frac{G^{\p}}{2G} = \sqrt{\frac{D}{2(D-1)}} ~\phi^{\p 2}.
\eea
Now we are ready to get the limiting behavior of $G$
by putting eq. (\ref{limit}) into eq. (\ref{G}) and integrating over $y$
\bea
\sqrt{G} & \sim & \int dy \frac{1}{(y_c - y)^{2\alpha}}.
\eea
All the sub-leading corrections are omitted
in the above expressions.

\subsection{$\alpha > 1$}

If $\alpha > 1$, then $\phi$, $\phi^{\p}$ and $G$ goes to $\infty$
as $y$ goes to $y_c$.
Integrating the previous equation,
\bea
\sqrt{G} & \sim & \frac{1}{(y_c-y)^{2\alpha-1}} \to \infty. \nn
\eea
The exponent of the warp factor also goes to $-\infty$ as
$y$ goes to $y_c$
\bea
\lim_{y\to \infty} -\int_0^{y} d\bar{y} \sqrt{G} & \sim & 
\lim_{y\to \infty} -\frac{1}{(y_c-y)^{2\alpha-2}} 
\rightarrow -\infty.
\eea
From the information on $\phi^{\p}$ and $G$, we can construct $U$
\bea
\label{yc}
U_{\rm leading} & = & G - \frac{\phi^{\p 2}}{2} \\ \nn
& = & c_1 \frac{1}{(y_c-y)^{4\alpha-2}}
- c_2 \frac{1}{(y_c-y)^{2\alpha}},
\eea
where $c_i$ with $i=1,2$ are positive constants.
Since we know the limiting behavior of $\phi$ itself,
we can reconstruct $U$ as a function of $\phi$ as $\phi$ goes to $\infty$.
\bea
\label{leading}
U_{\rm leading} & = & \bar{c}_1 {(\frac{\phi}{A})}^{n+2} - \bar{c}_2
{(\frac{\phi}{A})}^{n}, 
\eea
where $n = \frac{2\alpha}{\alpha-1} > 2$ is positive
and $\bar{c}_i$ are also positive constants.
Thus in the limit of $\phi \to \pm \infty$,
the first term dominates and determine the asymptotic form of the potential as
\bea
U_{\rm leading} & = & \bar{c}_1 {(\frac{\phi}{A})}^{n+2}. 
\eea
For even $n$, $A^{-n}$ is positive and $U$ is bounded from below
as $\phi \rightarrow \pm \infty$.
For odd $n$, we have to consider two cases.
First, $A > 0$. $U$ is bounded from below as $\phi \rightarrow \infty$.
Second, $A < 0$. We start from some $\phi_0$ and as $y \to y_c$,
$\phi \rightarrow -\infty$ and $U$ is also bounded from below.
This observation is very crucial. For odd $n$, the entire shape of $U$
for $\phi$ is not bounded from below. However, as long as the solution
is concerned, the $U$ is bounded from below.
Therefore, when $\alpha >1$, for both $A>0$ and $A<0$,
the potential $V$ is bounded from above in the solution.
For $\alpha > 1$, in the limit of $y_c - y \to 0^+$,
the first term dominates in eq. (\ref{yc})
and $U$ is always bounded from below.
Thus the potential $V$ is alway bounded from below
for $\alpha > 1$.

The $D$ dimensional energy density for $\alpha > 1$ is
\bea
\lim_{y \to y_c} 
\int_0^{y} d\bar{y} e^{-\int_0^{\bar{y}} d\tilde{y}
\sqrt{\frac{2D}{(D-1)}~G}} ~U,
\eea
and is finite since
$\lim_{x\rightarrow \infty} e^{-x^{2\alpha-2}} x^{4\alpha-2}$ is finite
where $x \sim \frac{1}{y_c-y}$.

We can conclude that for $\alpha > 1$
the potential $V$ is bounded from above in the solution
and $D$ dimensional energy density is finite.

\subsection{$\alpha = 1$}

This marginal case is very interesting
because the borderline of the criterion lies here.
The limiting behavior of the scalar field near the singularity is
\bea
\phi^{\p} & = & \frac{A}{y_c-y} \\ \nn
\phi & = & - A \log (y_c - y),
\eea
where the sub-leading terms are omitted.
From the equation \ref{friedmann}, we get 
\bea
\sqrt{G} &=& \sqrt{\frac{D}{2(D-1)}} A^2 \int dy \frac{1}{(y_c-y)^2} \\ \nn
& = & \sqrt{\frac{D}{2(D-1)}} A^2 \frac{1}{(y_c-y)} + {\rm nonsingular~~part}.
\eea
The next step is to get $U$ from $G$ and $\phi^{\p}$ from eq. (\ref{U})
\bea
U & = & G - \frac{\phi^{\p 2}}{2} \\ \nn
& = & \frac{1}{2} (\frac{D ~A^2}{D-1} - 1) \frac{A^2}{(y_c-y)^2}.
\eea

Since we know the limiting behavior of $U$ and $\phi$,
we can reformulate the leading term of $U(\phi)$ as
\bea
U(\phi)_{\rm leading} 
& = & \frac{1}{2} A^2 (\frac{D ~A^2}{D-1} -1) e^{\frac{2}{A} \phi}
= \frac{A^2}{2} \zeta e^{\frac{2}{A} \phi},
\eea
where
\bea
\zeta & = & \frac{D ~A^2}{D-1} -1.
\eea
We can extract the condition for $U$ bounded from below.
For $\zeta > 0$ ( $|A| > \sqrt{\frac{D-1}{D}}$), 
$U$ is bounded from below (the potential $V$ is bounded from above).
For $\zeta < 0$ ( $|A| < \sqrt{\frac{D-1}{D}}$), 
$U$ is not bounded from below (the potential $V$
is not bounded from above).

To evaluate $D$ dimensional energy density,
first we integrate $\sqrt{G}$.
The factor appearing in the exponent is
\bea
-\int_0^y dy^{\p} \sqrt{\frac{2D}{D-1}~G}
= \frac{D ~A^2}{D-1} \log (y_c-y) +{\rm finite ~~terms},
\eea
and $D$ dimensional energy density for $\zeta \neq 0$ is
\bea
\lim_{y \to y_c}
\int_0^{y} d\bar{y} (y_c-\bar{y})^{\frac{DA^2}{D-1}} U
& = & \lim_{y \to y_c} 
\int_0^{y} d\bar{y} \frac{A^2}{2} \zeta (y_c-\bar{y})^{\zeta-1} \\ \nn
& = & \lim_{y \to y_c} -\frac{A^2}{2} (y_c - y)^{\zeta}
+ {\rm finite~~ terms}.
\eea
For $\zeta > 0$, $D$ dimensional energy density is finite.
For $\zeta < 0$, $D$ dimensional energy density diverges.
We can further confirm the relation between two conditions.

It should be stressed that our criterion is
different from the criterion to avoid
the timelike naked singularity.
The metric can be expressed explicitly near the singularity
\bea
(\frac{1}{y_c-y})^{\frac{2 A^2}{D-1}} (-dt^2 + d\vec{x}^2) + dy^2
\eea
which clearly shows that for $|A| \ge \sqrt{D-1}$,
the naked singularity is null singularity
and it takes infinite time $t$ to arrive at the singularity $y_c$.
For $\sqrt{D-1} > |A| > \sqrt{\frac{D-1}{D}}$
{\footnote{$AdS_5$ supergravity has been studied in \cite{gubser},
and the region agrees with it for $D=4$.
$\xi$ in \cite{gubser} corresponds to $1/A$ but $\sqrt{2}$
appears due to the use of different units for 5 dimensional Planck scale.}}
,
even if the solution satisfies the criterion,
we have a naked timelike singularity.
Though Cauchy problem appears ill-defined 
in the presence of the timelike singularity,
we cannot rule out this case.
Coulomb branch solution of $D=4$ in \cite{gubser} 
is a definite example which has a timelike naked singularity
but can be resolved without facing the pathological problems
(except the unphysical Coulomb branch which can be ruled out
also by our criterion).
The criterion given in this paper is a more refined constraint
and even some types of timelike naked singularities
are admissible according to our criterion.
We have to mentions that a clear physical explanation of why
we can have a sensible physical theory even if Cauchy problem
is ill-posed is missing now.

\subsection{$\alpha < 1$}

It is easy to show that $U$ is not bounded
from below and $D$ dimensional energy density diverges.
$\phi$ goes to zero as $\phi^{\p}$ goes to $\infty$ 
(or as $y \rightarrow y_c$).
Therefore, the potential $V$ is not bounded from above.
For $\alpha < 1$, from eq. (\ref{yc}) and (\ref{leading}),
the leading term in the limit of $y \to y_c$
is the second term,
\bea
U_{\rm leading} & = & 
- c_3 \frac{1}{(y_c-y)^{2\alpha}} \nn \\
& = & - \bar{c}_3 {(\frac{A}{\phi})}^{m}, 
\eea
where $m = \frac{2\alpha}{1-\alpha} > 2$ is positive
and $c_3$ and $\bar{c}_3$ are also positive constants.
As we approach $y_c$, the scalar field $\phi$
goes to zero and $U \to -\infty$.
This is clear from the first line since we take the limit
$y_c - y \to 0^+$.
Thus the potential $V$ goes to $\infty$ and is not bounded from above.
By setting $x \sim \frac{1}{y_c-y}$,
the "$D$ dimensional energy density" formula is the same
as for $\alpha > 1$ and diverges for $\alpha < 1$.
Thus we confirm that for $\alpha < 1$
out criterion is not satisfied and also the potential is not
bounded from above.

\section{Discussion}

The above observations can be summarized in the following ways.
The criterion given here
is fundamental since it is the necessary condition
to have a consistent physical theory which allows us:
to have a semi-classical expansion around the classical solution
and to have a theory satisfying the consistency condition
without invoking the infinite tension brane.
The first property applies generally for the geometry
without involving the brane (without cutting the AdS).

All the examples studied here shows that two conditions,
the finiteness of $D$ dim. energy density and 
the potential bounded from above in the solution,
are the equivalent.
If we consider scalar potential which is bounded from above,
$D$ dimensional energy density of the solution remains finite.
Also if we restrict our interests only on the solution whose $D$ dimensional
energy density is finite, the potential is bounded from above.
It is puzzling that the two conditions which apparently 
look entirely independent are equivalent in all cases studied here.
We can give partial answer to this puzzle.
The anti-frictional force generated by either the potential energy
or the initial velocity $\phi^{\prime}$ destabilizes the scalar field,
and there are two options:

1. $\phi \rightarrow \pm \infty$ and $V(\phi) \rightarrow \infty$

2. $\phi \rightarrow \pm \infty$ and $V(\phi) \rightarrow -\infty$.

In case 1, the Hubble parameter is given by the difference
of $\phi^{\prime 2}$ and $V$, 
$h \sim \sqrt{\frac{\phi^{\prime 2}}{2} - V} $,
where both go to infinity
but the difference can be less than that.
In case 2, the Hubble parameter is given by the addition
of $\phi^{\prime 2}$ and $|V|$, 
$h \sim \sqrt{\frac{\phi^{\prime 2}}{2} + |V|} $,
which is larger than $|V|$ itself.
This gives rapid suppression of the warp factor and
the effective energy density can remain finite even though
$|V| \rightarrow \infty$. 
For the potential bounded from above,
the $\phi^{\prime 2}$ and $|V|$ give additional contribution
to warp factor enabling us to suppress the effective on-shell
Lagrangian density.
It is not easy to expect similar things in case 1
to make the energy density finite
since the scalar field runs to infinity too fast
and there is no chance to have a finite effective on-shell
Lagrangian density by the suppression of the warp factor.
Though no rigorous proof is available at this moment which can guarantee
the equivalence between two conditions (finite energy density
condition and the bounded above potential condition),
all the examples considered in this paper show that two conditions
are equivalent.
In this sense.
the approach used in this paper gives one explanation
of why Gubser's conjecture is working.

We leave the rigorous proof confirming the relations among the various
criteria given in \cite{MN1,gubser,gellmann,cohen} and
the finiteness of $D$ dimensional energy density,
and the issue related to timelike naked singularity
allowed in our criterion as future works.

\section*{acknowledgements}
Discussions with Prof. Kiwoon Choi were very helpful.
Youngjai Kiem encouraged me to write a paper
and gave lots of helps in refining the statements.
Without his aid this work would have not been finished.
Last part of this work has been done while I was visiting
Sungkyunkwan University.
I appreciate the hospitality of SKKU theory group during my visit.
I also thank D. B. Kaplan and C. Nunez who kindly 
informed me of other criteria on the singularities.
This work is supported by BK21 project
of the Ministry of Education and KRF Grant No. 2000-015-DP0080.

\newpage
\section*{table}
\begin{center}
Analogy of inflation models (de Sitter) and models in Anti-de Sitter space\\
in D dimensional spacetime ($t,\vec{x},y$) \\
\vspace{5mm}
\begin{tabular}{|c|c|c|}
\hline
& & \\
                        & Inflation/Deflation   
                        & AdS decreasing/increasing\\ 
                        & & \\ \hline
                        & & \\
~~Evolving parameter~~  & $t$           & $y$ \\ 
& & \\ \hline
                        & & \\
Potential       & $V(\phi(t))$          & $-V(\phi(y))$ \\ 
& & \\ \hline
& & \\
Metric  & $-dt^2 + e^{2 \int dt H_t }(d{\vec{x}}^2+dy^2)$       
& $e^{2 \int dy H_y }(-dt^2+d{\vec{x}}^2) + dy^2$  \\ 
& & \\ \hline
& & \\ 
Hubble parameter        
& $H_t = \pm\sqrt{\frac{2(\frac{\dot{\phi}^2}{2}+V)}{(D-1)(D-2)}}$      
& $H_y= \mp\sqrt{\frac{2(\frac{\phi^{\prime 2}}{2}-V)}{(D-1)(D-2)}}$ \\ 
& & \\ \hline
& & \\
Equation                
&~~ $\ddot{\phi}+ (D-1) H \dot{\phi} + \frac{dV}{d\phi} = 0$~~                  
&~~ $\phi^{\prime\prime}+ (D-1) H \phi^\prime - \frac{dV}{d\phi} = 0$~~
\\ 
& & \\ \hline
& & \\
Attractor condition & $V_{\rm min} > 0$ , $H_t > 0$ 
& $V_{\rm max} < 0$ , $H_y > 0$ \\ 
& & \\ \hline
& & \\
Einstein action & Exponential hierarchy & Exponential hierarchy \\ 
& & \\
Brans-Dicke action & Power law hierarchy & Power law hierarchy \\ 
& & \\ \hline
\end{tabular}
\end{center}

\end{document}